# A ZEROCROSSING ANALYSIS

## Emanuel Gluskin

Elec. Eng. Dept., Ben-Gurion Univ., Beer-Sheva 84105, and The Acad. Technol. Inst., Holon 58102, Israel.
Fax: 03-5026643; email: gluskin@ee.bgu.ac.il

We consider a direct representation of a *T*-periodic time function $f(t)$ by means of its zero-crossings (z-cs), $t_{k(s)} (\text{mod } T)$, $k, s = 1, 2, \ldots, m$, with even $m$. The use of the z-cs (simple isolated zeros) as the describing parameters is made possible by a singular model of a strongly nonlinear characteristic of an electrical element. This new method is found helpful in the calculation and synthesis of a practically important system, and deserves attention of the mathematicians.

Term *'z-c function'* denotes a continuous function with zero average, having only z-cs as its zeros. '∼' denotes direct proportionality. $\Omega$ denotes the frequency variable. $t_k (\text{mod } T)$ are the z-cs of $f(t)$, which belong to a *T*-interval. This interval is chosen so that at $t_1$ the slope of $f(t)$ is positive, thus sign $\frac{df}{dt}(t_k^+) \equiv \text{sign } \frac{df}{dt}(t_k^-) = (-1)^{k+1}$. $S \equiv \{t_k\}$. '$\langle \cdot \rangle$' is the integral average (over the period or the whole *t*-axis). $Z = Z(\Omega)$ is *impedance* [1]. The abbreviation 'z-c' is used here for both noun and adjective. $n \in \mathbf{N}$.

## The equation

Consider the equation
$$f(t) = \varphi(t) - \mu \mathbf{L}(\text{sign} f(t)), \qquad (1)$$

where $\varphi(t)$ is a *known T*-periodic z-c function. $\mu \geq 0$ is parameter. $\mathbf{L}$ is some first-order smoothing (integral) operator, generally with oscillatory kernel $G(\cdot)$: $(\mathbf{L}\zeta)(t) \equiv (G * \zeta)(t)$, on $(-\infty, t]$, for an integrable $\zeta(\cdot)$, which describes the steady-state non-resonant response of a linear circuit to the $\zeta(t)$. *If the function on which $\mathbf{L}$ acts possesses a nonzero average, $\mathbf{L}$ is required to provide a zero average of the resulting functions, i.e. $G * 1 = 0$.*

The requirement of the smoothing nature of $\mathbf{L}$ is that $(\frac{d}{dt}(\mathbf{L}\zeta))_\Omega \sim \zeta_\Omega/L_o$, as $\Omega \to \infty$, with some $L_o > 0$. As a realistic additional requirement (Section 5),

$(\frac{d}{dt}\mathbf{L} - 1/L_o)_\Omega = O(\Omega^{-d})$, $d \geq 2$, i.e., $\frac{d}{dt}\mathbf{L} - 1/L_o$ is a strongly smoothing operator. We also require that $\langle \zeta \mathbf{L} \zeta \rangle = 0$, for any periodic $\zeta(t)$, which means the simplifying assumption that by itself $\mathbf{L}$ describes a linear *lossless* (of a zero average power) subsystem.

Eq.(1) is inevitably reduced to an algebraic equation, which makes this integral equation and the role of $\mathbf{L}$ unique.

The z-cs of $\varphi(t)$ are denoted as $t_k^o$, $k = 1, 2, \ldots, m_o$, $S_o \equiv \{t_k^o\}$. We consider $\mu$ as a continuous parameter, $\mu \geq 0$, and the dependence of $t_k$ on $\mu$ means the mapping $S_o \to S(\mu)$.

The basic lemma is *that for an interval $0 \leq \mu < \mu_c$ $f(t)$ possesses the z-c features of $\varphi(t)$*, which means that though with a continuous increase in $\mu$, starting from zero and up to some critical $\mu_c$, the waveform of $f(t)$ is changed, no z-cs appear or disappear, i.e. $m(\mu) \equiv m(0) = m_o$. $\mathbf{L}$ provides continuity of the functions $\{t_k(\mu)\}$ in $[0, \mu_c)$, $t_k(0^+) = t_k^o$, i.e. $t_k$ are 'evolving' from $t_k^o$, and $\text{sign} f(t)$ in the interval $(t_k, t_{k+1})$ equals $\text{sign} \varphi(t)$ in the interval $(t_k^o, t_{k+1}^o)$.

We shall prove that with the increase in $\mu$, new zeros appear. Thus $\mu_c < \infty$ exists and is the limit for the range of the *z-c structural stability* [2-4].

$t_k$ may be unchanged (unmoved), i.e. $t_k(\mu) \equiv t_k^o$ ($S = S_o$) in $[0, \mu_c)$. In this case, $\text{sign} f(t) \equiv \text{sign} \varphi(t)$, and (1) is *solved*, $f(t) = \varphi(t) - \mu \mathbf{L}(\text{sign} \varphi(t))$. The 'constancy' of $t_k$ may be provided by proper synthesis of the operator $\mathbf{L}$. This, and also the importance of the 'constancy' to the *power* features of a practical circuit, will be seen after turning to a *z-c representation of $f(t)$*, which is our constructive point.

## The z-c representation of *f(t)*

For the z-c $f(t)$, $\text{sign} f(t)$ is the *rectangular-wave* function to be expressed as an explicit function of $t$ and $t_k$. Integrating $m$ mutually shifted and inverted combs of $\delta$-functions, we obtain

$$\text{sign} f(t) = D + \frac{2}{\pi} \sum_{k=1}^{m} (-1)^{k+1} \sum_{n=1}^{\infty} \frac{\sin n\omega(t - t_k)}{n}, \quad \omega = 2\pi/T, \qquad (2)$$

with a constant $D$. If $D$ is nonzero, then $\mathbf{L}$ is taken such that $\mathbf{L}(1) = 0$.
In view of (2), $\mathbf{L} \text{sign} f(t)$ depends on $t - t_k$. Thus (1) becomes:

$$f(t) = \varphi(t) - \mu F(t, \{t_s\}) \qquad (3)$$

with a *known* function $F$ of the $m + 1$ arguments $t, t_1, t_2, \ldots, t_m$, arranged in the $m$ differences $t - t_s$, i.e. $f(t)$ is presented using its z-cs.

The equation $f(t_k) = 0$ obtains the constructive form

$$\varphi(t_k) = \mu F(t_k, \{t_s\}). \qquad (4)$$

In accordance with the statements below, this equation has solutions. It is approximately solvable by a linearization, and in some cases (see example in Section 5) precisely. *Finding $t_k$ from (4), we obtain the solution of (1) in the form (3).*
In the case of $t_k$ independent of $\mu$, ($t_k = t_k^o$; $\varphi(t_k^o) = 0$, $\forall k$), eq. (4) becomes

$$F(t_k^o, \{t_s^o\}) \equiv (\mathbf{L}(\mathrm{sign}\,\varphi))(t_k^o) = 0, \quad \text{or} \quad F(t_k, \{t_s\}) \equiv (\mathbf{L}(\mathrm{sign}\,f))(t_k) = 0, \; \forall k. \quad (5)$$

Eq.(5) means that in the case of the 'constancy' of $t_k$, the set of the z-cs of $f(t)$, i.e. of the jump points of $\mathrm{sign}\,f(t)$, is included in the set of the *zeros* (not all of which are z-cs, see Fig.2 below) of $(\mathbf{L}\,\mathrm{sign}\,f)(t)$. (5) is also a condition on the $m-1$ (for $s \neq k$) among the $m$ parameters $t_k - t_s$, $s = 1, 2, .., m$, and also $T$. (Note that $t_{k+1} - t_k$, $k = 1, 2, .., m-1$, give any $t_k - t_s$, and $t_{m+1} = t_1 + T$.)

Formulated in terms of $\mathbf{L}$, eq.(5) is used below for a circuit synthesis.

For non-constant $t_k$, *with the precision* $O(\mu)$, eq.(4) yields

$$t_k \approx t_k^o + \mu[(d\varphi/dt)(t_k^o)]^{-1} F(t_k^o, \{t_s^o\}), \quad \forall k, \quad (6)$$

to be substituted in (3).

The critical parameter $\mu_c$ is found, together with the new (doubled) zero, from the equations $f(t) = 0$ and $df/dt = 0$,

$$\varphi(t) = \mu_c F(t, \{t_s(\mu_c)\}), \quad d\varphi(t)/dt = \mu_c \partial F(t, \{t_s(\mu_c)\})/\partial t. \quad (7)$$

Eqs.(7) show that *the limits for the z-c stability, or the z-c representation (3) of $f(t)$, can be found using only this representation.* These equations are strongly simplified in the case of constancy of $t_k$, when $\mu_c$ falls from $F$.

## Some supporting statements

**Lemma** 1 : *For any arbitrary T-periodic function $\psi(t)$, $\mu > 0$ exists such that for given $M > 0$ and $\mathbf{L}$, $\mu|(d(\mathbf{L}\,\mathrm{sign}\,\psi)/dt)(t_k^o)| < M$.*

*Proof*: Since (see Section 5) as $\Omega \to \infty$, $(\frac{d}{dt}\mathbf{L}\zeta - \zeta/L_o)_\Omega \sim \Omega^{-d}\zeta_\Omega$, $d \geq 2$, then writing $|\frac{d}{dt}\mathbf{L}\zeta| \leq |\frac{d}{dt}\mathbf{L}\zeta - \zeta/L_o| + |\zeta|/L_o$ with $\zeta = \mathrm{sign}\,\psi$ ($\zeta_\Omega \sim 1/\Omega; |\zeta| = 1$), we have $|\zeta|/L_o = 1/L_o$, and
$|\frac{d}{dt}\mathbf{L}\zeta - \zeta/L_o| \sim |\sum(\pm n)^{-d}\zeta_{\pm n\omega}\,e^{\pm int}| \leq \sum|(\pm n)^{-d}\zeta_{\pm n\omega}| \leq (\sum n^{-2d})^{1/2}(\sum|\zeta_{\pm n\omega}|^2)^{1/2} = (\sum n^{-2d})^{1/2}$
The required $\mu$ may thus be easily chosen. The independence of $\mu$ on the specific form of $\psi$ is important for Theorem 1 below.

**Corollary** 1 : *$\mu > 0$ may be found such that for any certain admissable $\psi$, $\varphi(t)$ and $\mathbf{L}$,*
$$\mu|(d(\mathbf{L}\,\mathrm{sign}\,\psi)/dt)(t_k^o)| < \min_k \{|(d\varphi/dt)(t_k^o)|\}.$$

**Corollary** 2 : *For a small $\mu$, and given $\varphi(t)$, the addition of $-\mu\mathbf{L}(\mathrm{sign}\,\psi(t))$ to $\varphi(t)$ (see (1)) can not cause any new z-c, i.e. $f$ possesses the z-c features of $\varphi$.*

Indeed, since $\mathbf{L}\,\mathrm{sign}\,\psi$ is limited, we can find $\mu$ so small that $\mu|\mathbf{L}\,\mathrm{sign}\,\psi|$ would be smaller than $|\varphi(t)|$ at any extreme point of $\varphi(t)$ (there is a finite number of such points in

the period), and for $\mu$ also satisfying the condition of Corollary 1, the addition of $-\mu\mathbf{L}(\text{sign}\,\psi(t))$ cannot cause any new zero, obviously.

**Theorem** 1 : *For $\mu$ defined by Corollary 2, eq.(1) has a T-periodic z-c solution with $m = m_o$.*

*Proof*: The iterations $f_o = \varphi$, $f_n = \varphi - \mu\mathbf{L}(\text{sign}\,f_{n-1})$, $n = 1, 2, \ldots$ create T-periodic functions. According to Corollary 2, a *certain* $\mu > 0$ may be found such that *each $f_n$* will be a z-c function, possessing $m_o$ z-cs per period. Consider, for such $\mu$, the infinite number of the sets $S$, given by all of the $f_n$, belonging to a certain period, as the $(\infty \times m_o)$-matrix of all of the z-cs. The infinite set $(\infty \times 1)$ (the first *column* of the matrix) of all of the *first* z-cs ('$t_1$' for each $f_n$) has a point of condensation in this finite interval (the period), and similarly for the other $m - 1$ infinite sets of the '$t_2$', ..., '$t_m$' z-cs of $\{f_n\}$. This yields the existence of a *T*-periodic $f(t)$ having the z-c features of $\varphi(t)$, which is a solution of (1).

**Theorem 2**: *This z-c solution is unique.*

*Proof* (by contradiction): Since $\langle\zeta\mathbf{L}\zeta\rangle = 0$, the function $(\zeta\mathbf{L}\zeta)(t)$ alternates its polarity. Assume that (1) has two different solutions, $f_1$ and $f_2$. Then $f_1 - f_2 = -\mu\mathbf{L}(\text{sign}\,f_1 - \text{sign}\,f_2)$. Multiplying by $(\text{sign}\,f_1 - \text{sign}\,f_2)$, we obtain an alternative-polarity function in the right-hand side, and the nonnegative expression $(|f_1|+|f_2|)[1 - (\text{sign}\,f_1)(\text{sign}\,f_2)]$ in the left-hand side.

**Remark:** By multiplying by $\text{sign}\,f$, we similarly prove that the equation $f(t) = \mu\mathbf{L}(\text{sign}\,f)$ possesses only a zero solution. Since with an unlimited increase in $\mu$, (1) becomes the latter equation, this conclusion means that with an unlimited increase in $\mu$, $f$ cannot keep the z-c features of $\varphi$, i.e. $\mu_c < \infty$. Considering Fig.2 below, one sees, however, that a time shift in one of the sides of $f(t) = \mu\mathbf{L}(\text{sign}\,f)$ may lead to a nonzero solution.

Introducing our circuit, we show now that optimization of an important functional of $f(t)$ (eq.(9) below) is obtained for $S(\mu) \equiv S_o$, i.e. for constant z-cs.

## The electrical circuit

The theory of the strongly nonlinear *fluorescent lamp circuits* is an important application of the z-c analysis, and the main circuit parameter, power, is subject to a specific optimization, treated here in terms of the z-cs.

The lamp's *v-i* (*voltage-current*) characteristic (considered in detail in [5,6]) is close to $v(i) = A\,\text{sign}(i)$, where '$A$' is a constant ($\approx 115$ volt). This 'hardlimiter' model introduces the z-cs of the lamp's current function into the circuit equation that is of the type (1).

The lamp's circuit is shown in Fig.1. Because of the hardlimiter-type characteristic, directly applying the sinusoidal line voltage to the lamp would cause an unlimited current, and a 'ballast' (sub)circuit B is needed. We consider B to be *linear time-invariant*, and, for simplification of some formulae, *lossless* circuit. It may thus include only inductors ($L$) and capacitors ($C$). At least one inductor (the electrical analog of mass, for series connection of the elements) must be present for limiting $i(t)$. The inductor(s) also smooth the waveform of $i(t)$, which provides the structural stability of the z-cs of $i(t)$, with respect to a range of the input parameter $U$ that strongly influences $i(t)$ and the circuit operation.

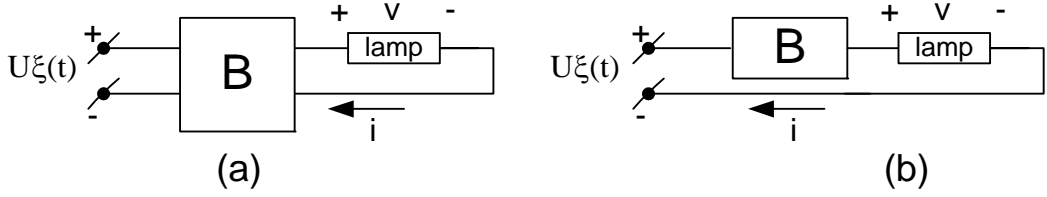

Figure 1: (a): $i(t)$ is the output current of te 2-port (the two-operator version of B), $v = A\,sign(i)$. (b): the 1-port version ($L_1 = L_2$ in eq.(8)).

$U$ (in volts) is the amplitude (scaling factor) of the input voltage function $U\xi(t)$, $|\xi(t)| \le 1$. $U\xi(t)$ is required here to be periodic and such that number of the z-cs per period of the response function $i(t)$ is finite. The integral character of the linear operators $\mathbf{L}_1$ and $\mathbf{L}_2$ in the circuit equation (8) is inevitably caused by the inductor(s) in B. An important requirement of both $\xi(t)$ and B is that no resonant process in B occurs, i.e. the frequency spectrum of $\xi(t)$ is completely different from the spectra of the operators that describe B. Despite these limitations, we have significant freedom in the structure of B, for a synthesis of this circuit, required by applications.

For the circuit shown in Fig.1(a), the linearity of B results, via the superposition with respect to the voltages on the ports of B, in:

$$i(t) = U(\mathbf{L}_1\xi)(t) - A\mathbf{L}_2(sign[i(t)]). \qquad (8)$$

Eq.(8) is reduced to (1) by denoting $(\mathbf{L}_1\xi)(t)$ as $\varphi(t)$, $\mathbf{L}_2$ as $\mathbf{L}$, $A/U$ as $\mu$, $i(t)/U$ as $f(t)$, and by using that $sign[i(t)] = sign[i(t)/U]$.

Regarding the frequency spectrum of $i(t)$, we see from (8) that for a sinusoidal $\xi(t)$, the high harmonics in $i(t)$ are independent of $U$.

Using the z-cs of $i(t)$ as an analytical tool, we now derive some important conclusions regarding the lamp's power.

**The lamp's power and its sensitivity to the variations in $U$.**

By the definition of the (average) power, $P = \langle iv \rangle = \langle i\,A\,sign\,i \rangle = A\langle |i| \rangle = AU\langle |f| \rangle$, and using (8) and then that $\langle \zeta \mathbf{L} \zeta \rangle = 0$, we also obtain

$$P = \langle [U\varphi(t) - A\mathbf{L}(sign\,i(t))]A\,sign\,i \rangle = AU\langle \varphi(t)\,sign f(t) \rangle. \qquad (9)$$

Since $t_k(\mu)$ are included in $sign f$,

$$P = AU\eta(\mu)$$

where $\eta = \langle \varphi \, \text{sign} f \rangle = \langle |f| \rangle > 0$.

In practice a very important parameter (function) of such nonlinear circuits is the *sensitivity of P to changes in U, expressed in the relative values*:

$$K_U(\mu) \equiv (dP/P)/(dU/U) = d\ln P/d\ln U = 1 + d\ln \eta/d\ln U.$$

Since $d\ln \mu/d\ln U = -1$, we have

$$K_U(\mu) = 1 - d\ln \eta/d\ln \mu = 1 - (\mu/\eta)d\eta/d\mu. \qquad (10)$$

**Theorem** 3 : $K_U \geq 1$, and $K_U(\mu) \equiv 1$ iff $t_k$ are constant.

*Proof*: For $\mu \in [0, \mu_c)$

$$\frac{d}{d\mu}(\text{sign } f)(t) = -2\sum_1^m \text{sign}[(df/dt)(t_k)]\delta(t - t_k(\mu))dt_k/d\mu, \qquad (11)$$

and since $\text{sign}[(df/dt)(t_k)] = \text{sign}[(d\varphi/dt)(t_k^o)]$, (10) yields

$$K_U(\mu) = 1 - (\mu/\eta)d\langle \varphi \, \text{sign} f\rangle/d\mu = 1 - (\mu/\eta)\langle \varphi d\,\text{sign} f/d\mu\rangle$$

$$= 1 + 2\mu(\eta T)^{-1}\sum_1^m \text{sign}[(d\varphi/dt)(t_k^o)]\varphi(t_k)dt_k/d\mu.$$

For a small $\mu$, $\varphi(t_k) \sim ((d\varphi/dt)(t_k^o))(t_k - t_k^o)$, and $dt_k/d\mu \sim (t_k - t_k^o)/\mu$; thus

$$K_U(\mu) = 1 + 2(\eta T)^{-1}\sum_1^m |(d\varphi/dt)(t_k^o)|(t_k(\mu) - t_k^o)^2, \qquad (12)$$

and it is obvious that $K_U(\mu) \geq 1$ and the minimal (the lowest possible curve) $K_U(\mu) \equiv 1$ is obtained iff $t_k(\mu) \equiv t_k^o$. □

For non-constant $t_k$, and $\mu$ close to $\mu_c$, the power sensitivity may possess unacceptably high values. This depends (see (6)), first of all, on $dt_k/d\mu \sim F(t_k^o, \{t_s^o\}) = (\mathbf{L}(\text{sign}\,\varphi))(t_k^o)$. The range 1÷5 for $K_U$ may be easily obtained in the lamp circuits for not very strong variations in $U$. Contrary to that, if $P$ were the power of a resistor *in any linear system*, then $P \sim U^2$, and $d\ln P/d\ln U \equiv 2, \forall U$.

We see that $K_U$ well characterizes the circuit's nonlinearity, and since in (9) $\varphi$ depends on $\mathbf{L}$, the role of the linear elements (or $\mathbf{L}$) in $K_U$ is unusual ([5,6] for more details).

Minimization of $K_U$ is an important topic in the theory of the fluorescent lamp circuits.

## Example of the synthesis of B that minimizes $K_U$

For the common case of $\varphi(t) = -\varphi(t+T/2)$, *with 2 z-cs per period*, we obtain $f(t) = -f(t+T/2)$, and $t_2 - t_1 = T/2 = t_2^o - t_1^o$.

Using the *susceptence* [1], $B(\Omega) \equiv \text{Im}[(\mathbf{L}e^{i\Omega t})/(e^{i\Omega t})]$ (this is simply $\text{Im}[Z(\Omega)^{-1}]$), $B_n \equiv B(n\omega)$, and Fourier series expansion, we turn the equality

$$F(t_k, \{t_s\}) \equiv (\mathbf{L}\,\text{sign}f)(t_k) = 0, \quad k = 1, 2, \ldots, m, \quad (13)$$

into

$$F(t_1, \{t_1, t_2\}) = \frac{4}{\pi} \sum_1^\infty \frac{B_{2n-1}}{2n-1} = 0. \quad (14)$$

Since in (14) $B_{2n-1}$ cannot be all of the same sign, circuit B must be capacitive at certain frequencies and inductive at others. The simplest appropriate circuit, *inductive at high frequencies*, is the series *L-C* ballast. (Replace B in Fig.1(b) by such a circuit.) For such a B,

$$B_n = (\frac{1}{n\omega C} - n\omega L)^{-1}, \quad (15)$$

and (14) becomes

$$\frac{4}{\pi} \sum_1^\infty \frac{1}{(2n-1)^2 - (\frac{\omega_o}{\omega})^2} \equiv -\frac{\omega}{\omega_o} \text{tg}(\frac{\pi\omega_o}{2\omega}) = 0, \quad \omega_o = (LC)^{-1/2}, \quad (16)$$

where $\omega_o$ is the resonant frequency of the *L-C* circuit.

Eq.(16) gives $\omega_o = 2\omega, 4\omega, \ldots$. In the practice of the circuits, $2\omega$, or even a somewhat smaller value should be used, in order not to increase too much (with respect to the simplest possible, purely inductive B), harmonic currents of the frequencies $3\omega, 5\omega, \ldots$.

The condition $\omega_o \lesssim 2\omega$ is only one of two possible conditions on the two elements $L$ and $C$. The other condition is that *P should be of the nominal value for a nominal value of U*.

The function $(\mathbf{L}\,\text{sign}f)(t)$ is shown, for $\omega_o = 2\omega$, in Fig.2. We see the role of $\omega_o$, and note that the set of z-cs of $(\mathbf{L}\,\text{sign}f)(t)$ cannot coincide with $S$. The latter can be proved *in the general case* by contradiction; $|\langle\text{sign}f\mathbf{L}\,\text{sign}f\rangle|$ would then be a large value, but $\mathbf{L}$ is such that $\langle\zeta\mathbf{L}\zeta\rangle = 0$.

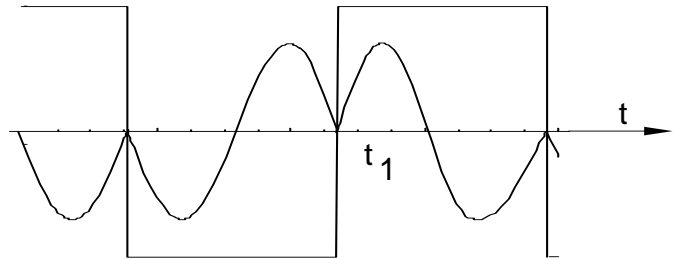

Figure 2: (*sign f*)(*t*) and (*L sign f*) for the L-C-lamp circuit, for $\omega_o = 2\omega$.

For a more complicated $\varphi(t)$, i.e. for $m_o = 4, 6, \ldots$, B would be required to include correspondingly more elements, in order to ensure $K_U \equiv 1$. For any $S_o$ a B (or **L**) can be synthesized, using not more that $m_o$ elements of the $L$ and $C$ types, such that the requirements for $K_U$ and $P$ are satisfied.

Expanding, in the general case, the relevant *polynomial fraction* $B_n$, describing B, into partial fractions of the type (15), we can represent B as a parallel connection of the simple $L - C$ series circuits, with different $\omega_o$. According to the expansion of (15) by the powers of $1/n$,

$$(\frac{1}{n\omega C} - n\omega L)^{-1} = -\frac{1}{n\omega L}[1 + (\frac{\omega_o}{n\omega})^2 + \ldots],$$

we have, in the general case, as $n \to \infty$ (or $\Omega \to \infty$)

$$B_n = -\frac{1}{n\omega L_o}[1 + O(n^{-2})], \text{ or } B(\Omega) = -\frac{1}{\Omega L_o}[1 + O(\Omega^{-2})].$$

If linear resistive elements, causing power losses, were included in B, then an $O(\Omega^{-2})$-term would replace $O(\Omega^{-3})$ in the more general equality $(\frac{d}{dt}\mathbf{L} - 1/L_o)_\Omega = O(\Omega^{-d})$, $d \geq 2$, which was used in Lemma 1.

## Final remarks

For the circuit with the series *L-C* ballast, the equation for $i(t)$ may be written as

$$L\frac{di}{dt} + A \operatorname{sign} i(t) + \frac{1}{C}\int_{-\infty}^{t} i(\tau)d\tau = U\xi(t). \qquad (17)$$

The nonlinearity of (17) is obvious when using sign($\cdot$), *but not when using* (sign $i$)($\cdot$) *given by series (2)*. The nonlinearity in (2) is due to the fact that the shifting parameters $t_k$ *belong to the unknown solution*. Following [7], we call such a "hidden" nonlinearity a 'z-c nonlinearity'. There are other important examples of this 'z-c nonlinearity' [7-12], forming a class of dynamic problems where the constructive role of the z-cs of *a priori* unknown functions, and the appearance of the z-cs as time-shifts, are typical. Thus, it is shown in [8] that a linear oscillatory equation (as (17) when $A = 0$); with a sinusoidal right-hand side, for a function $x(t)$, with the additional 'reflecting' condition $(dx/dt)(t_k^+) = -(dx/dt)(t_k^-)$, where $t_k$ are zeros of $x(t)$, may lead to a chaotic $x(t)$. Much earlier work [9] shows that a 'shift-nonlinearity', with the shift defined by the trajectory of a particle in cyclotron, leads to a typical *nonlinear* resonance. A relevant nonlinear analysis of also *stochastic* processes is suggested in [10] and discussed in [11]. In general, since z-cs are easily observable, detectable, and suitable for control parameters, the z-c nonlinearity is very important.

Finally, a standard 'dynamical systems' formulation may be given to our problem. Considering **L** as a solution operator for an LTI equation of the type

$$\sum_0^{r_{\max}} a_r \frac{d^{2r}}{dt^{2r}} y(t) = \zeta(t) ,$$

using the 'phase variables' $x_1 = y$, $x_2 = \dot{x}_1$, ..., $x_r = \dot{x}_{r-1}$ , we replace (1) by

$$\dot{\mathbf{x}} = \mathbf{f}(\mathbf{x}, t) \qquad (18)$$

with $\text{sign}\, x_2(t)$ included in the right-hand side.

Some systems with discontinuous right-hand sides have already been investigated, e.g. in the known works by N.N. Bautin on clock synchronization, and the known book by A.F. Filipov. However this is the first time that such a *practically important system* as the fluorescent lamp circuit (such circuits consume in total about 20% of all electrical power generated) has been analyzed using the discontinuity in (18).

## Acknowledgements

I am grateful to M. Lifsic (M.S. Lifshitz), S. Shnider, V. Gotlib, B-Z. Shklyar and the late S.G. Mikhlin for discussions.